\documentclass[a4paper]{jpconf}

\usepackage{graphicx}
\begin{document}
\title{Signatures of a quantum Griffiths phase in a d-metal alloy close to its ferromagnetic
quantum critical point}

\author{Almut Schroeder$^1$, Sara Ubaid-Kassis$^1$ and Thomas Vojta$^2$}

\address{$^1$ Department of Physics, Kent State University, Kent OH 44242, USA}
\address{$^2$ Department of Physics, Missouri University of Science and Technology, Rolla MO 65409, USA}

\ead{aschroe2@kent.edu}

\begin{abstract}
We report magnetization ($M$) measurements close to the ferromagnetic quantum phase transition of the
d-metal alloy Ni$_{1-x}$V$_x$ at a vanadium concentration of $x_c \approx11.4 \%$.
In the diluted regime ($x>x_c$), the temperature ($T$) and magnetic field  ($H$) dependencies of the magnetization
are characterized by nonuniversal power laws and display $H/T$ scaling in a wide
temperature and field range. The exponents vary strongly with $x$  and follow the predictions of a
quantum Griffiths phase.  We also discuss the deviations and limits of the quantum Griffiths phase as well
as the phase boundaries due to bulk and cluster physics.
\end{abstract}
%The temperature ($T$) dependence of the magnetic susceptibility is best described by simple nonuniversal power laws, $M/H(T,H \rightarrow 0) \sim T ^{-\gamma}$ , rather than Curie Weiss laws. Moreover, the magnetic field ($H$) dependence of the low-temperature magnetization displays power laws $M \sim H^{\alpha}$  with $\alpha=1-\gamma$. This leads to $H/T$ scaling of the magnetization in a wide temperature ($10K < T \leq 300K$) and field ($H \leq 5T$) range. The exponent $\gamma$ is strongly $x$ dependent, decreasing from 1 at $x \approx x_c$ to $\gamma < 0.1$ for $x=15\%$. This behavior clearly differs from both classical and quantum critical behavior in a clean 3D ferromagnet. Instead, it closely follows the predictions for a quantum Griffiths phase associated with a quantum phase transition in a disordered metal. \end{abstract}

%\pacs{71.27.+a, 75.40.-s, 75.50.Cc}

\section{Introduction}

Magnetic quantum phase transitions (QPT) have been studied in transition metal
alloys and in heavy-fermion compounds tuned, e.g., by pressure or chemical
substitution. They still offer challenges to theory and experiment
(see Ref.\ \cite{hvl07} for a recent review). Quantum critical behavior is
signified by singularities in thermodynamic and transport
properties. Usually, specific power laws with characteristic exponents have
been predicted at the quantum critical point (QCP) for ``clean'' homogeneous systems,
while ``disordered'' inhomogeneous systems, driven, e.g., by chemical substitution,
may show different behavior \cite{vojta06}.
%Recent theories address the impact of disorder on QPTs more systematicallyF
%and new universality classes have been proposed
%(for a review, see Ref.\ \cite{vojta06}).
In the case of metallic (itinerant) Heisenberg magnets, a
strong-disorder renormalization group  \cite{vojta0709} predicts an
exotic infinite-randomness QCP  accompanied by quantum Griffiths singularities
\cite{vojta05}. At such a QCP, thermodynamic observables are expected
to be singular not just at criticality but in a finite region around the QCP called the
quantum Griffiths phase (GP). This region features power laws
(e.g., in the magnetic susceptibility,  $\chi \sim T^{\lambda-1}$, and the magnetization, $M\sim H^\lambda$)
characterized by a nonuniversal Griffiths exponent $\lambda$ which varies with distance to the QCP.
Quantum Griffiths singularities have attracted a lot of attention. Many heavy fermion compounds
display anomalous power-laws in specific heat $C(T)$ and $\chi(T)$  \cite{stewart01};
and quantum Griffiths behavior was suggested as an explanation \cite{castro00}.
Recently, a more systematic variation of the exponents  could be
found at the ferromagnetic QPT of CePd$_{1-x}$Rh$_{x}$ \cite{westerkamp09CePd}.\\
%According to the standard theory of ferromagnetic quantum criticality in 3D
%metals \cite{HertzMillis}, specific heat $C$, magnetic susceptibility $\chi$ and
%electrical resistivity $\rho$ should behave as $\chi \sim T^{-4/3}$,  $C/T \sim \log(T)$
%and  $\rho(T) \sim T^{5/3}$ when approaching the QCP at low temperatures $T$. This was
%observed in Ni$_{x}$Pd$_{1-x}$ with $x=0.025$, at least over a limited temperature regime
%\cite{nicklas99nipd}.
%Many ferromagnetic binary alloys such as Ni$_{1-x}$Cu$_x$ or Ni$_{1-x}$V$_x$
%in which $T_c$ can be tuned by chemical substitution $x$ show a still more complicated
%behavior, even in the paramagnetic phase.
%In early investigations \cite{amamou75rho}, the existence of large magnetic clusters with
%giant local moments was proposed to describe the magnetization $M$ data of these inhomogeneous
%systems.

To avoid additional complications due to the Kondo effect and to study a larger energy scale
we recently investigated the simple fcc transition metal alloy  Ni$_{1-x}$V$_x$  \cite{boelling68}
as an example of an itinerant ferromagnet (FM) in which the transition temperature ($T_c=630K$ for pure Ni)
can be tuned to zero by chemical substitution. As explained by Friedel \cite{friedel58}
the ``disorder'' is introduced because the charge contrast of the replacing vanadium atoms
%a V impurity (with
%5 fewer electrons than Ni) creates a localized charge and a spin reduction on the
%neighboring Ni-sites. This reduces not only the average spin moment but
%also
creates large defects yielding an inhomogeneous magnetization density.
In contrast, diluting Ni  with isoelectronic Pd does not lead to a strongly disordered scenario: Ni$_{1-x}$Pd$_{x}$ remains ferromagnetic up to $x_c=0.975$ where it rather shows the signatures of a clean quantum critical point \cite{nicklas99nipd}.
We showed in Ref.\ \cite{ubaid2010} that magnetization and susceptibility above the
critical vanadium concentration $x_c\approx 11.4\%$ where $T_c$ is suppressed to 0 indeed follow simple power laws with nonuniversal exponents that confirm the quantum Griffiths scenario over a wide
temperature and magnetic field regime. At very low temperatures, deviations from the quantum
Griffiths scenario hint at a cluster glass phase. Here, we provide additional details not shown in Ref.\
\cite{ubaid2010}.  We demonstrate that $H/T$-scaling holds for a wide concentration regime and show the
scaling plots. In addition, we reveal how the impact of disorder is manifest in the original $M$ data close
to $x_c$, and we show the details of the determination of the phase boundaries in order to better distinguish
bulk behavior from individual cluster physics in this inhomogeneous system.

\begin{figure}[h]
\includegraphics[width=18pc]{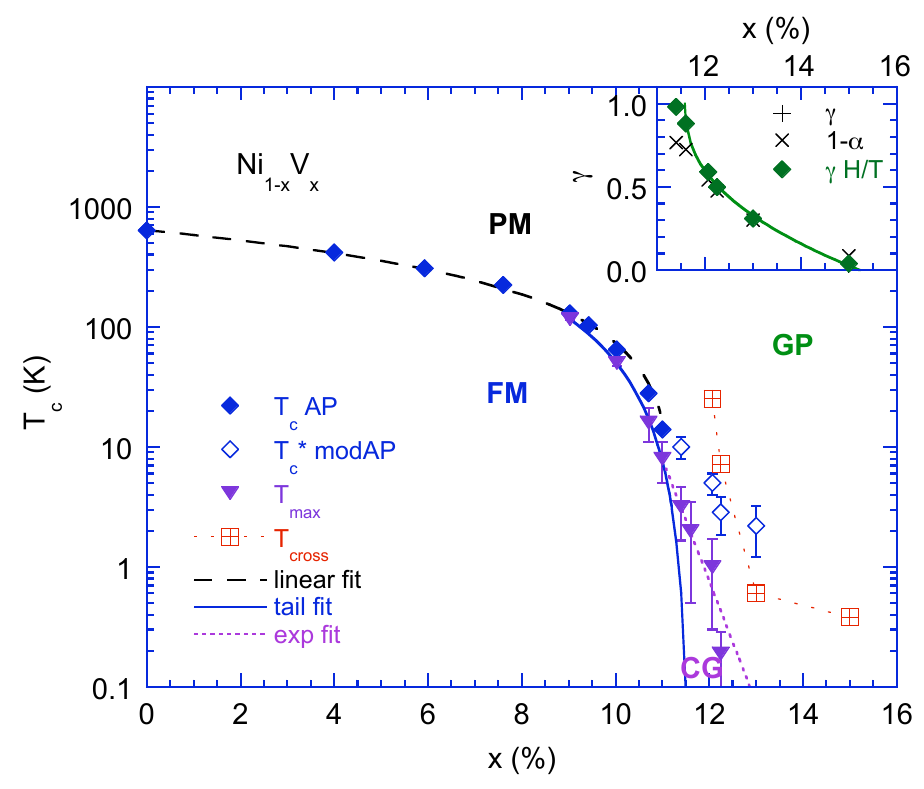}\hspace{2pc}%
\begin{minipage}[b]{18pc}\caption{\label{asfiga3}Temperature ($T$) - concentration ($x$) phase diagram of Ni$_{1-x}$V$_x$ showing the ferromagnetic (FM), paramagnetic (PM), quantum Griffiths (GP) and cluster glass (CG) phases. Closed and open diamonds mark $T_c$ and $T_c*$ as determined by linear and modified Arrott plots, respectively (data from [8] included). Triangles denote to $T_{max}$  from maxima in susceptibility. $T_{cross}$ is seen as lower limit of GP (see \cite{ubaid2010}). The dashed line is a linear extrapolation, the solid line is a tail fit  describing the FM boundary, while the straight dotted line marks the onset of the CG. Inset shows the strong concentration $x$-variation of the exponent $\gamma$ from $\chi(T)$, $\alpha - 1$  from $M(H)$ and $\gamma$ from the $H/T$- scaling plot in the GP.}
\end{minipage}
\end{figure}

\section{Results}

 Magnetization and ac-susceptibility measurements were performed on polycrystalline  Ni$_{1-x}$V$_x$ samples with $x=9-15\%$ as described in Ref.\ \cite{ubaid2010}.
An orbital contribution of $\chi_{orb}=6\times 10^{-5} emu/mol$ has been subtracted from all data shown
($M_m=M-\chi_{orb} H$).\\

Figure 1 shows the temperature-concentration phase diagram. For $x\leq11\%$, the critical temperature $T_c$ was determined by the standard Arrott analysis. Plots of
$M^2$ vs. $H/M$ as in Fig.\ 2  show straight parallel isotherms which implies $M^2 = M_0^2(T) +c H/M$ as is common for itinerant magnets (in Fig.\ 2, only low 
$T$ data are shown).  $T_c$ is then extracted via the mean-field $T$-dependencies of  $M_0(T)$ and susceptibility $(-c/M_0^2(T))$ \cite{ubaid08}.
The resulting $T_c(x)$
can be simply extrapolated linearly (dashed line) from the high $T_c=630K$ of nickel down to 0 at $x\approx 11\%$ \cite{boelling68}.

For  $x\geq11\%$, the straightforward AP analysis does not longer work because the data in Fig.\ 2 are not described by straight
lines. Introducing ``exponents'' as in a classical critical regime leads to a ``modified" Arrott plot \cite{arrott67} implying the behavior
$ M^{1/\beta} = M_0^{1/\beta}(T) +c (H/M)^{1/\gamma}$. A good description for $x>11\%$ of the $M(H>0.5T,T)$ data in a wide regime (outside any critical regime) can be achieved with $\beta=0.5$ and $\gamma(x)<1$ \cite{ubaid08}, as indicated by the dotted fit line in Fig.\ 2.
The resulting transition temperatures $T_c*$ of these modified Arrott plots remain finite up to $x=15\%$, while other extrapolations in Fig.\ 2 using smaller $H/M$ values would lead to smaller $T_c$.

\begin{figure}[h]
\begin{minipage}{18pc}
\includegraphics[width=17pc]{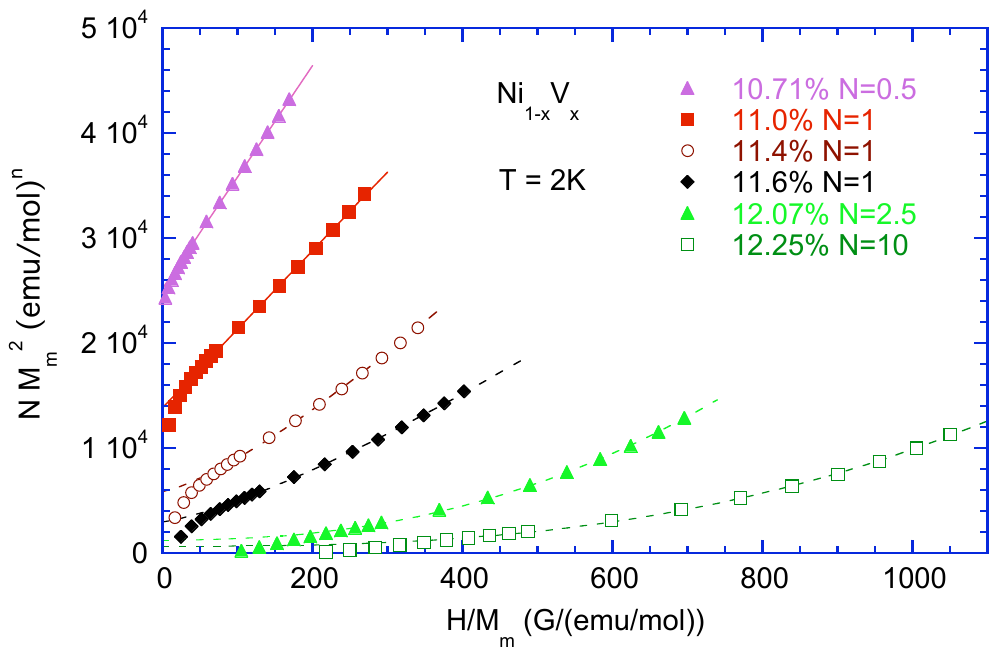}
\caption{\label{asfiga1}Probing Arrott plots ($M^2-M_0^2 \sim (H/M)^{1/\gamma}$ with $\gamma=1$) for various V-concentrations $x$. Far clarity only low temperature $T=2K$ data (with modified y-value) are shown. For $x>11\%$ a modified Arrott plot (with  $\gamma (x) < 1$) is a good description of the data above $H \approx 0.5T$ as indicated by the dashed line.}
\end{minipage}\hspace{2pc}%
\begin{minipage}{18pc}
\includegraphics[width=18pc]{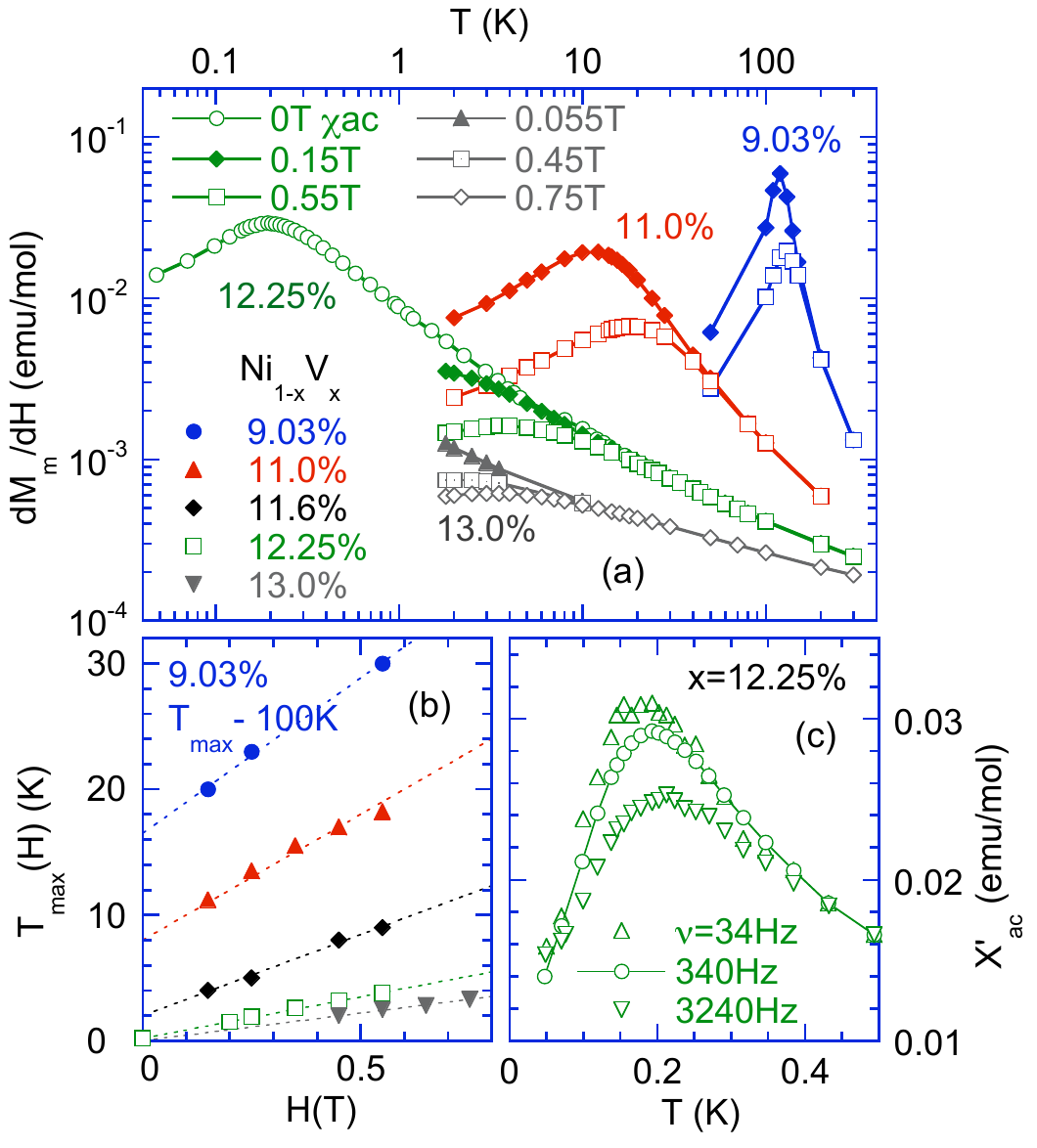}
\caption{\label{asfiga2a}
Determination of transition temperature $T_{max}$.
(a) Differential susceptibility $dM_m/dH$ vs temperature $T$ in small magnetic fields $H$ (symbol)  displaying maxima at $T_{max}(H)$ for different $x$.  (b) $T_{max}(H)$ vs magnetic field $H$. The dashed line indicates the linear extrapolation to determine $T_{max}=T_{max}(H \to 0)$. (c) ac-susceptibility $\chi'_{ac}$ vs. temperature $T$ in $H=0$T with $H_{ac}=0.1$G at different frequencies $\nu$ for $x=12.25\%$.}
\end{minipage}
\end{figure}
In addition to Arrott
plots, we analyze  field-dependent maxima at $T_{max}(H)$ in the differential
susceptibility $\chi(T) = dM(T)/dH$
indicating spin ordering or freezing as shown in Fig.\ 3(a). Fig.\ 3(b) shows the  linear
extrapolation of $T_{max}(H)$ taken at 0.55T to 0.1T to determine $T_{max}=T_{max}(H \to 0)$.
%$T_{max}(x)$ is somewhat lower than $T_c$
%derived from the high field mean field analysis. It remains finite for $x<13\%$ (see Fig.1).
In particular for $x=12.25\%$ a frequency dependent maximum at $T_{max}=0.19K$ was determined by $\chi_{ac}(T,\nu=380Hz)$ in zero field with $H_{ac}=0.1G$ which hints at the onset of a cluster glass \cite{ubaid2010}. $T_{max}$ increases by 0.018K per decade in frequency \cite{ubaid2010} as shown in Fig.3(c).
Although a detailed study of the evolution with dilution $x$ of the cluster growth and dynamics is still outstanding,
we can already note the qualitative effects of disorder on the ferromagnetic ordered state  for $x>11\%$.
As is obvious in Fig.\ 1, the high and low field extrapolation lead to {\it different} transition temperatures     ($T_c, T_c* > T_{max}$) hinting at  cluster
freezing for $x>11\%$.
%We emphasize that deviations from linear Arrott plots and the sensitivity of $T_c$
%towards the extrapolation procedure already point to an unconventional QPT.
The $x$-dependence of $T_{max}$ in the accessible temperature region is better
described by an exponential (dotted line) rather than a power law. Also,
a ``tail'' fit to ($\ln(T/T_0)\sim (x_c-x)^{-\nu\psi}$, see \cite{vojtalt}) rather than a power law serves as a good description of the onset of FM order for data
between about $9\%$ and  $11\%$ leading to $x_c \approx 11.6\%$ (solid line). The discrepancies between the various methods
and the spin-glass like features at the lowest temperatures suggest that the real QCP is masked at very
low $T$ by ordering of clusters.\\

Nonetheless, at sufficiently high temperatures (in the region $T_{max} < T < T_c(0\%)$) cluster ordering does not seem to play a role,
and various quantities display power laws.
Figs.\ 4(a) and (b) present the $H$ and $T$ dependencies of the magnetization as $M/H$ for various $x$.  Fig.\ 4(b) shows essentially the susceptibility $\chi$,
since $\chi=M/H=dM/dH$ for low fields  ($H < 0.5T$) and high $T$ ($T>20K$, $T > T_c$).
While for $x\leq11\%$ the negative slope in the log-log plot $\gamma=-dln(\chi_m)/dln(T)$ increases with falling $T$ towards $T_c$,  for $x>11\%$, $\chi(T)$
follows a simple power law for $20K<T<300K$. The exponent decreases from $\gamma(x=11.4\%)=1$  to $\gamma(x=15\%)= 0.04$.
$M/H(H)$ follows a power law  $M/H \sim H^{\alpha-1}$ for high $H$. For $x<11\%$, where $M(H)$ nearly saturates, the exponent $1-\alpha$ is close to  1, and
therefore very different than $\gamma$. However, for $x>12\%$, the high-field exponent $1-\alpha$ matches the susceptibility
exponent $\gamma$. The deviations from a power law at low fields in Fig.\ 4(a) are due to the finite $T$ limitations.

\begin{figure}[h]
\begin{minipage}{18pc}
\includegraphics[width=17pc]{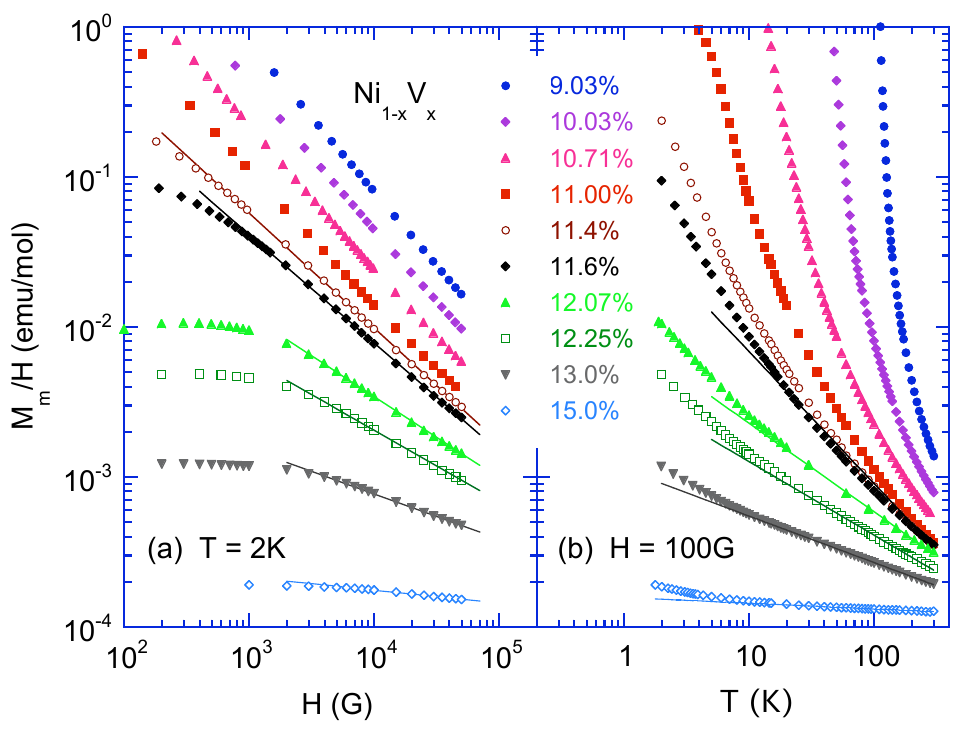}
\caption{\label{asfiga4} (a) Magnetic field $H$ and (b) temperature $T$ dependence of the magnetization $M$ for a wide $x$ regime. $M/H(H,T)$ follows a power law (solid line) with the same exponent  ($\alpha -1$) in (a) as  ($\gamma$) in (b) for all $x>11.6\%$. }
\end{minipage}\hspace{2pc}%
\begin{minipage}{18pc}
\includegraphics[width=18pc]{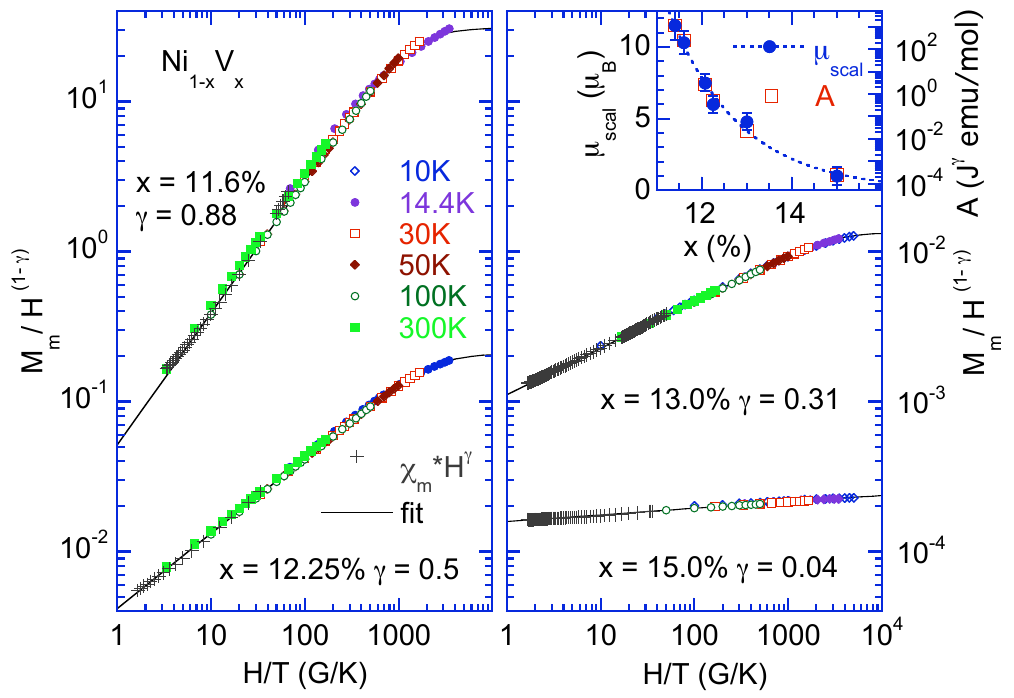}
\caption{\label{asfiga5}$H/T$ scaling plot showing some $M(H,T)$ data within $10K - 300K$ and $100G - 50 kG $ for different $x$. The line represents a fit using $Y(z)$ (see text). Inset shows fit parameters $\mu_{scal}$ and $A$ vs $x$. }
\end{minipage}
\end{figure}

Since both the $M(T)$ and $M(H)$ show power laws with the same exponent, simple $H/T$ scaling is expected for $x>12\%$. Fig.\ 5 shows the
scaling plot using the form
 $M/H = H^{-\gamma} \, Y(\mu_{scal} H/k_B T)$
where $Y$ is the scaling function and $\mu_{scal}$ is a scaling moment for several $x$.
All $M(H,T)$ data for $T\geq14 K$ collapse,
confirming $H/T$ scaling.  The scaling function $Y$ is well approximated by
the form
$Y(z)=A'/(1+z^{-2})^{\gamma/2}$ where
$A'=A/\mu^{\gamma}$ is a constant.
This phenomenological form arises from simply combining the two limiting power laws with the same exponent $\gamma$ in the $H-T$ plane,
$(M/H)^{-1} = H^{\gamma} Y^{-1} \sim [(\mu_{scal} H)^2+(k_B T)^2]^{\gamma/2}$.
Close to $x=11.6\%$,  the quality of the collapse is less satisfactory. The resulting exponent $\gamma$ (which
matches that obtained by a fit of $\chi(T)$ for all $x$
between 11.4\% and 15\%) is shown in the inset of Fig.\ 1. The scaling moment $\mu_{scal}$ and amplitude $A$ are shown in the inset of Fig.\ 5,
demonstrating the growth of the typical cluster
size and number with $x \to x_c$.

The consistent power laws, and in particular, the $H/T$ scaling of $M(H,T)$ are in excellent agreement with the predictions
for a quantum Griffiths phase with Griffiths
exponent  $\lambda=\alpha = 1-\gamma$. A critical concentration of $x_c=11.4\%$ can be identified from the condition
$\gamma(x_c)=1$ (neglecting logarithmic terms).
Fitting to power law $1-\gamma(x) \sim (x-x_c)^{\nu \psi}$ as predicted by theory \cite{vojta0709}
yields $x_c=11.6\%$ with $\nu\psi=0.42$ as shown in the upper
inset in Fig.\ 1. This value is in close agreement with the ``tail'' fit of $T_{max}(x)$.\\

\section{Conclusions}

On the one hand, our results confirm that Ni$_{1-x}$V$_x$ follows the scenario of an infinite-randomness QCP
with a quantum Griffiths phase, as expected in an itinerant Heisenberg magnet \cite{vojta0709,vojta05}.
The QCP at $x_c\approx11.6\%$ has been estimated by extrapolations from outside
the critical region, where the cluster ordering is less disturbing (through $\gamma(x_c)\to 1$ and $T_{max}(x_c)\to0$).
On the other hand, we see clear signs of cluster ordering towards $x_c$, in particular deviations from scaling
at lower temperatures (such as the upturns in Fig.\ 4(a) as well as model dependent transition temperatures for $x>11\%$.
As discussed in Ref.\ \cite{ubaid2010}, the magnetization $M(H,T>T_{max})$ for $x>12\%$  can be well described by an additional ``Curie term'' due
to frozen clusters which exceeds the term due to the fluctuating (Griffiths) clusters below $T_{cross}$ (see Fig. 1). Such a change in low-temperature
behavior was predicted to occur in itinerant Heisenberg systems due to the RKKY interactions \cite{dobro05}.
A Griffiths phase with nonuniversal power laws at higher $T$ (but below $T_c(0\%)$) combined with a cluster glass (CG) (indicated by maxima in $\chi(T)$) at very low $T$ has also been
observed in other diluted compounds (CePd$_{1-x}$Rh$_{x}$ \cite{westerkamp09CePd}, URu$_{2-x}$Re$_{x}$Si$_2$\cite{bauer05})
close to a ferromagnetic transition with much lower $T_c$
and can be understood as a generic feature of this disordered itinerant QPT \cite{vojtalt}.\\

This work has been supported in part by the NSF under grant nos. DMR-0306766, DMR-0339147,
and DMR-0906566 and by Research Corporation.

%\bibliography{nivref}{}
%\bibliographystyle{iopart-num}

%\end{document}

\end{document}